\documentstyle[12pt,epsfig]{article}
\newcommand{\be}{\begin{equation}}

\newcommand{\en}{\end{equation}}
\newcommand{\beeq}{\begin{eqnarray}}
\newcommand{\eneq}{\end{eqnarray}}

\makeatletter
\def\vereq#1#2{\lower3pt\vbox{\baselineskip1.5pt \lineskip1.5pt
\ialign{$\m@th#1\hfill##\hfil$\crcr#2\crcr\sim\crcr}}}
\makeatother
\begin{document}
\begin{titlepage}
\begin{flushright}
TUM-HEP-643/06
\end{flushright}
\begin{center}
{\Large\bf 
A universal seesaw mechanism
 in five dimensions
} 
\end{center}
\vspace{1cm}
\begin{center}
Naoyuki {Haba}
\footnote{E-mail: haba@ph.tum.de 
}
\end{center}
\vspace{0.2cm}
\begin{center}
{\it Physik-Department, 
 Technische Universit$\ddot{a}$t M$\ddot{u}$nchen,
 James-Franck-Strasse,
 D-85748 Garching, Germany}
\\[0.2cm]
{\it Institute of Theoretical Physics, University of
Tokushima, 
770-8502 Tokushima, Japan}\\

\end{center}
\vspace{1cm}
\begin{abstract}

We show the universal seesaw
 in the extra dimension setup, where 
 three extra 
 vector-like fields exist in the 5D bulk
 with heavy masses. 
We take 
 the framework of the left-right symmetric model. 
The universal seesaw formula is easily obtained as
 a replacement of the vector-like mass in 4D case $M_i$ to
 $2 M_* \tan [\pi RM_i]$ ($M_*$: 5D Planck scale,
 $M_i$: vector-like bulk mass, and 
 $R$: compactification radius). 
The smallness of Dirac neutrino mass 
 can be naturally explained in the 5D setup. 
We also show the Majorana neutrino case.

\end{abstract}
\end{titlepage}
\newpage
\section{Introduction}

The origin of quark and lepton masses is 
 still a mystery, which can be a very important
 key for searching a fundamental theory 
 beyond the standard model (SM). 
Why are fermion masses except
 for top quark mass 
 much smaller than electroweak breaking scale,
 and why are neutrino masses much smaller than 
 other fermion masses?  
The ordinal seesaw mechanism was suggested to explain 
 the smallness of neutrino masses\cite{seesaw}. 
The smallness of quark and lepton masses (except for
 top quark mass) 
 comparing to the electroweak breaking
 scale can be explained in a different framework,  
 where vector-like extra fermions 
 are introduced with heavy masses. 
This mechanism is so-called universal seesaw 
 mechanism\cite{useesaw}-\cite{SU5}. 
The extra fermions have vector-like masses 
 so that they have no effects to the 
 electroweak precision measurements. 
The left-right symmetric
 model with a gauge group, 
 $SU(3)_c \times SU(2)_L\times SU(2)_R \times U(1)_{B-L}$,
 is one example which can have a universal
 seesaw structure. 
For the top quark mass which is the same
 order as electroweak symmetry breaking,
 we should take special setup\cite{tanimoto}\footnote{Another
 approach for the large top mass has been studied in
 $S_3$ flavor symmetry\cite{useesaw2.3}.}.

In this paper 
 we show the universal seesaw mechanism 
 in the flat extra dimension setup\cite{Arkani-Hamed:1998rs}. 
We show the 5D universal seesaw 
 in the framework of the left-right symmetric model. 
We introduce  
 extra 
 vector-like ($SU(2)_L\times SU(2)_R$ singlet) fields 
 in the 5D bulk, and other fermions on the 4D brane. 
The 5D universal seesaw formula is easily obtained as
 a replacement of the vector-like mass in 4D case $M_i$ to
 $2 M_* \tan [\pi RM_i]$ ($M_*$: 5D Planck scale,
 $M_i$: vector-like bulk mass, and 
 $R$: compactification radius)\cite{haba}. 
As for the neutrinos, 
 the smallness of Dirac neutrino masses 
 can be naturally explained in the 5D setup. 
Only Dirac neutrino masses can be very small 
 when the bulk neutrino masses are close to a half of
 compactification scale. 
We will also show the Majorana neutrino case, where
 5D seesaw mechanism works. 
For the heavy top quark mass, 
 the vector-like extra ``top'' quarks should be
 localized on the 4D brane.

This paper is organized as follows. 
In section 2, we show a brief review 
 of 4D universal seesaw. 
Section 3 shows the 5D universal seesaw
 formula. 
For the neutrino sector, 
 we will study both Dirac and Majorana
 neutrino cases. 
Section 4 shows the summary.

\section{4D universal seesaw mechanism}

At first we briefly review
 the 4D universal seesaw mechanism in 
 the left-right symmetric model,
 which has a gauge group, 
 $SU(3)_c \times SU(2)_L\times SU(2)_R \times U(1)_{B-L}$.
The quark sector is given by
\begin{eqnarray}
& Q_L{}_i = 
\left(
\begin{array}{c}
u_i \\
d_i
\end{array}
\right)_L =(3,2,1,1/3), \;\;\;\;\;\;
& Q_R{}_i = 
\left(
\begin{array}{c}
u_i \\
d_i
\end{array}
\right)_R=(3,1,2,1/3), \\
&U_L{}_{i} = (3,1,1,4/3), \;\;\;\;\;\; 
&U_R{}_{i} = (3,1,1,4/3), \\
&D_L{}_{i} = (3,1,1,-2/3), \;\;\;\;\;\; 
&D_R{}_{i} = (3,1,1,-2/3). 
\end{eqnarray}
The Higgs sector are 
\begin{equation}
H_L=(1,2,1,1), \;\;\;\;\; H_R=(1,1,2,1),
\end{equation}
which take vacuum expectation values (VEVs) as
\begin{equation}
\langle H_L \rangle =
\left(
\begin{array}{c}
0 \\
v_L
\end{array}
\right) =
(0,v_L,0,0), \;\;\;\;\; 
\langle H_R \rangle =
\left(
\begin{array}{c}
0 \\
v_R
\end{array}
\right) = (0,0,v_R,0),  
\end{equation}
where 
 $ v_R \gg v_L$ 
 must be satisfied with  
$v_R \geq {\cal O}(10^4)$ GeV from
 the experimental search of $W_R$\cite{PDG}.

The most general Yukawa interactions 
 in the quark sector are
\begin{eqnarray}
\label{MU}
{\cal L}_{mass}^q &=& 
 y^{u,L}_{ij} \overline{Q_L}_i \tilde{H}_L U_{Rj} +
 y^{u,R}_{ij} \overline{Q_R}_i \tilde{H}_R U_{Lj} +
 M_{\cal U}{}_{ij} \overline{U_L}_i U_{Rj} + {\rm h.c.} \\
&+& y^{d,L}_{ij} \overline{Q_L}_i {H}_L D_{Rj} +
 y^{d,R}_{ij} \overline{Q_R}_i {H}_R D_{Lj} +
 M_{\cal D}{}_{ij} \overline{D_L}_i D_{Rj} + {\rm h.c.}, 
\end{eqnarray}
where $\tilde{H}=\epsilon H^*$ and
 $M$'s are bare mass parameters.
Without loss of generality, 
 $M_{\cal U}{}_{ij}$ ($M_{\cal D}{}_{ij}$)
 can be a real diagonal
 matrix, ${\rm diag}(M_U,M_C,M_T)$ (${\rm diag}(M_D,M_S,M_B)$),
 through a bi-unitary transformation. 

Then, 
 the up- and down-type mass matrices are
 given by 
\begin{eqnarray}
(\bar{u}, \bar{U})_L
\left(
\begin{array}{cc}
 0 & y^{u,L} v_L^* \\
y^{u,R}{}^\dagger v_R & M_{\cal U} \\
\end{array}
\right)
\left(
\begin{array}{c}
u \\
U \\
\end{array}
\right)_R, \;\;\;
(\bar{d}, \bar{D})_L
\left(
\begin{array}{cc}
 0 & y^{d,L} v_L \\
y^{d,R}{}^\dagger v_R^* & M_{\cal D} \\
\end{array}
\right)
\left(
\begin{array}{c}
d \\
D \\
\end{array}
\right)_R, 
\end{eqnarray}
respectively. 
We take real VEVs, for simplicity, and
 note 
 $m^u_L{}_{ij}=y^{u,L}_{ij} v_L$,
 $m^u_R{}_{ij}=y^{u,R}_{ij} v_R$,
 ${m}^d_L{}_{ij}={y}^{d,L}_{ij} v_L$, and
 ${m}^d_R{}_{ij}={y}^{d,R}_{ij} v_R$. 
Since we have already used the degrees of
 freedom for 
 $M_{\cal U,D}$
 to be a flavor diagonal,
 field redefinitions remain only in $Q_{L,R}$. 
These field redefinitions can make 
 $m_{L,R}$ 
 triangular matrices as
\begin{equation}
m_{L,R}= 
\left(
\begin{array}{ccc}
\times & 0 & 0 \\
\times & \times & 0 \\
\times & \times & \times \\
\end{array}
\right),
\label{triangular}
\end{equation}
where
 $\times$ stands for non-zero element\cite{tanimoto}.

The $|m^{u,d}_{L}|$ is of order 100 GeV and
 $|m^{u,d}_{R}|\geq 10^4$ GeV. 
When 
 $|M_{\cal U,D}| \gg |m^{u,d}_{R}|$, 
 light three generation quarks' masses 
 are given by 
\begin{eqnarray}
&&(M^u_{light})_{ij} \simeq 
    - (m^u_L)_{ik} \;{1 \over M_{\cal U}{}_k}\; (m^u_R{}^T)_{kj}, 
     \;
  (M^d_{light})_{ik} \simeq 
    - (m^d_L)_{ik} \;{1 \over M_{\cal D}{}_k}\; (m^d_R{}^T)_{kj},
\label{u10}
\end{eqnarray}
which is so-called universal seesaw
 formula. 
We assume $m^u_L$ and $m^u_R$ 
 are diagonal matrices of the flavor base from now on. 
We do not make any particular
 assumption on the structure
 of the down-type Yukawa 
 matrix, which is 
 the origin of flavor mixings in 
 the quark sector.

Considering the bottom quark mass, 
 the scale of $|M_B|$ is expected to be of order 
 $10^2 \times |m^{b}_{R}|$. 
When $|M_B|\sim |M_S|\sim |M_D| \sim |M_T| \sim |M_C|\sim |M_U|$, 
 other quark masses can be reproduced through 
 the universal seesaw with the mass hierarchies
 of $m_{L,R}$\footnote{It is possible to reproduce
 quark and lepton mass hierarchies by the mass
 hierarchies of $M_{\cal U,D}$, but  
 in this paper we take a standing point
 that vector-like masses are all the same order.} 
 except for 
 the top quark mass. 
In order to obtain 
 the suitable magnitude of top
 quark mass, 
 $|m_R^t|$ should be the same order as 
 (or even larger than) 
 $|M_T|$\cite{tanimoto}. 
In this case, the top quark and heavy extra $T, \overline{T}$ 
 quark masses
 are given by 
\begin{eqnarray}
m_{t} \simeq  
 {{m_L^t}{m_R^t}\over 
 \sqrt{m^{t2}_R+M^2_{T}}}, \;\;\;\;\;\;
m_{T} \simeq  \sqrt{m_R^{t2}+M^2_{T}},
\end{eqnarray}
respectively.
This suggests $m_t={\cal O}(10^2)$ GeV.

Next, let us show the 
 lepton sector, which is
 give by
\begin{eqnarray}
& L_L{}_i = 
\left(
\begin{array}{c}
\nu_i \\
e_i
\end{array}
\right)_R =(1,2,1,-1), \;\;\;\;\;\;
& L_R{}_i = 
\left(
\begin{array}{c}
\nu_i \\
e_i
\end{array}
\right)_R=(1,1,2,-1), \\
&N_L{}_{i} = (1,1,1,0), \;\;\;\;\;\; 
&N_R{}_{i} = (1,1,1,0), \\
&E_L{}_{i} = (1,1,1,-2), \;\;\;\;\;\; 
&E_R{}_{i} = (1,1,1,-2). 
\end{eqnarray}
The most general Yukawa interactions 
 in the lepton sector are given by
\begin{eqnarray}
{\cal L}_{mass}^l&=& {\cal L}_0 + {\cal L}' \nonumber \\
{\cal L}_0 &=& 
 y^{e,L}_{ij} \overline{L_L}_i {H}_L E_{Rj} +
 y^{e,R}_{ij} \overline{L_R}_i {H}_R E_{Lj} +
 M_{\cal E}{}_{ij} \overline{E_L}_i E_{Rj} + {\rm h.c.} \nonumber \\
&+& 
 y^{\nu,L}_{ij} \overline{L_L}_i \tilde{H}_L N_{Rj} +
 y^{\nu,R}_{ij} \overline{L_R}_i \tilde{H}_R N_{Lj} +
 M_{{\cal N}}{}_{ij} \overline{N_L}_i N_{Rj} + {\rm h.c.}
\label{150} \\
{\cal L}' &=& 
 \hat{y}^{\nu,L}_{ij} \overline{L_L}_i \tilde{H}_L N_{Lj}^c +
 \hat{y}^{\nu,R}_{ij} \overline{L_R}_i \tilde{H}_R N_{Rj}^c \nonumber \\
&+& 
 M_{{\cal N}L}{}_{ij} \overline{N^c_L}_i N_{Lj} +
 M_{{\cal N}R}{}_{ij} \overline{N^c_R}_i N_{Rj}  + {\rm h.c.}
\label{15}
\end{eqnarray}
${\cal L}_0$ conserves the lepton number while 
 ${\cal L}'$ breaks the lepton number. 

When ${\cal L}'=0$, the neutrino and 
 charged lepton mass matrices are 
 given by 
\begin{eqnarray}
(\bar{\nu}, \bar{N})_L
\left(
\begin{array}{cc}
 0 & y^{\nu,L} v_L^* \\
y^{\nu,R}{}^\dagger v_R & M_{\cal N} \\
\end{array}
\right)
\left(
\begin{array}{c}
\nu \\
N \\
\end{array}
\right)_R, \;\;\;
(\bar{e}, \bar{E})_L
\left(
\begin{array}{cc}
 0 & y^{e,L} v_L \\
y^{e,R}{}^\dagger v_R^* & M_{\cal E} \\
\end{array}
\right)
\left(
\begin{array}{c}
e \\
E \\
\end{array}
\right)_R,
\end{eqnarray}
which are similar to the quark mass matrices.  
We take real VEVs 
 and note 
 $m_{L,R}^\nu = y^{\nu, L,R} v_{L,R}$,  
 $m_{L,R}^e = y^{e, L,R} v_{L,R}$. 
We can expect $|m^{\nu,e}_L|={\cal O}(10^2)$ GeV and
 $|m^{\nu,e}_R| \geq {\cal O}(10^4)$ GeV. 
The masses of light three generation leptons 
 are given by 
\begin{eqnarray}
&&(M^\nu_{light})_{ij} \simeq 
    - (m^\nu_L)_{ik} \;{1 \over M_{\cal N}{}_k}\; (m^\nu_R{}^T)_{kj}, 
     \;
      (M^e_{light})_{ij} \simeq 
    - (m^e_L)_{ik} \;{1 \over M_{\cal E}{}_k}\; (m^e_R{}^T)_{kj} ,
\end{eqnarray}
when $|m^{\nu,e}_R| \ll |M_{\cal N,E}|$.  
Similar to the down-type quark sector, 
 tau mass suggests 
 $|M_\tau| \simeq 10^2 \times |m^{\tau}_R|$. 
The mass hierarchy of $m^{e}_{L,R}$ can
 realize the mass hierarchy of the
 three generation lepton masses 
 when $|M_E|\sim |M_\mu|\sim |M_\tau|$.  
For the suitable tiny Dirac neutrino masses, 
 $|M_{\cal N}|$ must be extremely larger 
 than other vector-like masses as 
 $|M_{\cal N}|\simeq 10^{12} \times |m^\nu_R|$.\footnote{
Another possibility is that 
 $y^\nu$s are extremely smaller than 
 other Yukawa couplings of quarks and 
 charged leptons.} 
It is just assumption and we should explain why
 $|M_{\cal N}|$ is so huge comparing to other
 vector-like masses.

When ${\cal L}'\neq 0$, the neutrino 
 mass matrix becomes  
\begin{eqnarray}
(\overline{\nu_L}, \overline{\nu^c_R}, \overline{N_L}, \overline{N^c_R})
\left(
\begin{array}{cccc}
0 & 0 & \hat{y}^{\nu,L} v_L^* & {y}^{\nu,L} v_L^* \\
0 & 0 & {y}^{\nu,R} v_R   & \hat{y}^{\nu,R} v_R \\
\hat{y}^{\nu,LT} v_L^* & y^{\nu,R}{}^T v_R & M_{{\cal N}L}  & M_{{\cal N}} \\
y^{\nu,L}{}^T v_L^*    
 &  \hat{y}^{\nu,RT} v_R 
 &  M_{{\cal N}}^T  &  M_{{\cal{N}}R} \\
\end{array}
\right)
\left(
\begin{array}{c}
{\nu_L^c} \\
{\nu_R} \\
{N_L^c} \\
{N_R}
\end{array}
\right),
\label{neu4d}
\end{eqnarray}
where $\psi_L^c \equiv C \psi_L^T$ and 
 $\hat{m}^\nu_{L,R} = \hat{y}^{\nu, L,R} v_{L,R}$.
Under the condition, 
$(|m^\nu_L|,  |\hat{m}^\nu_L| \ll \ ) \ 
 |m^\nu_R|, |\hat{m}^\nu_R| \ll |M_{{\cal N}}|, 
 |M_{{\cal N}L}|, |M_{{\cal N}R}|$,
the mass matrix for light six neutrinos 
 is given by
\begin{eqnarray}
\label{20}
M_{light}&\simeq&
-\left(
\begin{array}{cc}
\hat{m}_L^\nu & m_L^\nu \\
m_R^\nu & \hat{m}_R^\nu \\
\end{array}
\right)
\left(
\begin{array}{cc}
 M_{{\cal N}L} & M_{{\cal N}} \\
 M_{{\cal N}}^T  & M_{{\cal N}R} \\
\end{array}
\right)^{-1}
\left(
\begin{array}{cc}
\hat{m}_L^{\nu T} & m_R^{\nu T} \\
m_L^{\nu T} & \hat{m}_R^{\nu T} \\
\end{array}
\right).  
\label{20n}
\end{eqnarray}
This mass matrix has been analyzed\cite{useesaw1.5}. 
Here we show one example
 of the special limit, 
 $\hat{m}^\nu_{L,R}=0$, in which Eq.(\ref{20n}) 
 becomes 
\begin{eqnarray}
M_{light} &\sim&
\left(
\begin{array}{cc}
-m_L^\nu M^{-1}m_L^{\nu T} &
 m_L^\nu M'^{-1}m_R^{\nu T} \\
 m_R^\nu M'^{-1}m_L^{\nu T} &
-m_R^\nu M''^{-1}m_R^{\nu T} \\
\end{array}
\right),
\label{17}
\end{eqnarray}
where 
 $M$'s stands for
 functions of
 $M_{{\cal N}L}, M_{{\cal N}R}, M_{{\cal N}}$.\footnote{
 We assume non-zero determinant of
 the inverse 
 matrix of $M_{{\cal N}L}, M_{{\cal N}R}, M_{{\cal N}}$ 
 in
 Eq.(\ref{20}).}  
$|m_L^\nu | \ll |m_R^\nu |$ suggests that 
 the order of neutrino mass matrix of the 
 light three generation 
 is given by 
\begin{equation}
M_{light}^\nu \simeq -m_L^\nu M^{-1} m_L^{\nu T},
\label{23}
\end{equation}
which is the same as the usual seesaw mechanism. 
Notice that Eq.(\ref{23}) 
 holds independently of
 how larger $m_R^\nu$ is than 
 $m_L^\nu$ as long as $M$'s are the same order. 
When $M$'s are of order $10^{14}$ GeV,
 the suitable tiny neutrino masses 
 are obtained through the seesaw mechanism.

\section{5D universal seesaw mechanism}

Now we take 
 the 5D setup, 
 where the 5th dimension coordinate $(y)$
 is compactified on $S^1/Z_2$.
Only 
 $SU(2)_L \times SU(2)_R$ 
 singlets 
 are spread in the 5D bulk, while
 the other quarks and leptons are
 localized on the 4D brane. 
We can consider two scenarios for
 the neutrino sector. 
One is the Dirac neutrino case and the
 other is Majorana neutrino case. 
We will show two cases in the following
 two subsections.

\subsection{Dirac neutrino case}

This case has the same Yukawa interactions 
 for all quarks and leptons,
 which is corresponding to 4D case of ${\cal L}_0 =0$ 
 in Eq.(\ref{15}). 
Quarks and leptons are all Dirac fields. 
Here we show the neutrino mass matrix, 
 for an example. 
Other mass matrices of quarks and leptons
 are obtained in the same way.

Since the 5D theory is a vector-like
 theory, we must introduce 
 the chiral partners 
 $N^m$'s for $N$'s, 
 which are mass dimension 3/2. 
Under $Z_2$ parity, $y \rightarrow -y$,
 ${N}_{L}\equiv(N_{l},N^m_{l})^T$
 (${N}_{R}\equiv(N_{r}^m,N_{r})^T$)
 transforms as
 ${N}_{L}(x^\mu, -y)= \gamma^5 {N}_{L}(x^\mu, y)$
 (${N}_{R}(x^\mu, -y)= -\gamma^5 {N}_{R}(x^\mu, y)$).
The mode expansions are given by
\begin{eqnarray}
&& {N}_{L}(x^\mu,y) \;=\; \frac{1}{\sqrt{\pi R}} \left(
  \begin{array}{r}
    \frac{1}{\sqrt{2}}\,N_{l}^{(0)}(x^\mu)+
    \sum\limits_{n=1}^\infty
    \cos (\frac{ny}{R})\,N_{l}^{(n)}(x^\mu) \\
    \sum\limits_{n=1}^\infty
    \sin (\frac{ny}{R})\,{N^m_{l}}^{(n)}(x^\mu)
  \end{array} \right) ,
\label{mode11} \\
&& {N}_{R}(x^\mu,y) \;=\; \frac{1}{\sqrt{\pi R}} \left(
  \begin{array}{r}
    \sum\limits_{n=1}^\infty
    \sin (\frac{ny}{R})\,N^m_{r}{}^{(n)}(x^\mu) \\
    \frac{1}{\sqrt{2}}\,{N_{r}}^{(0)}(x^\mu)+
    \sum\limits_{n=1}^\infty
    \cos (\frac{ny}{R})\,{N_{r}}^{(n)}(x^\mu) 
  \end{array} \right) ,
\end{eqnarray}
where the factor $1/\sqrt{2}$ of the zero-mode
 is needed for the canonical
 kinetic term in the 4D effective Lagrangian. 
$N_{l}$, $N_r^m$ ($N_{r}$, $N_l^m$)
 are left- (right-)handed Weyl fermions, 
 in which $N_{l}$ ($N_{r}$) 
 has a 
 zero-mode and survive below the
 compactification scale, $R^{-1}$.

The bulk Dirac mass of the vector-like neutrinos 
 is given by 
\begin{equation}
\label{3.8}
  {\cal L}_{\rm 5DM} \;=\;
  M_{\cal N}{}_{ij} \overline{N_{Li}}N_{Rj} + {\rm h.c.} ,
\end{equation}
which is invariant under the
 $Z_2$ parity, and conserves the 
 lepton number. 
We can always take the 
 diagonal base of the
 generation index for $M_{\cal N}{}_{ij}$. 
We assume their eigenvalues are real
 for simplicity in the following discussions. 
By integrating out the 5th dimension, 
 the 
 4D effective Lagrangian is obtained, where 
 Eq.(\ref{3.8}) becomes
\begin{eqnarray}
 {\cal L}_{\rm 4DM}\, \;&=&\; 
   \sum\limits_{n=0}^\infty M_{\cal N}\;
  \overline{N_{l}}^{(n)}{N_{r}}^{(n)}
+ \sum\limits_{n=1}^\infty M_{\cal N}\;
  \overline{N_{l}^m}^{(n)}{N_{r}^m}^{(n)} 
 +{\rm h.c.}.
\end{eqnarray}
We omit spinor indices here. 
Kaluza-Klein (KK) masses are given by
\begin{eqnarray}
&&  {\cal L}_{\rm KK} \;=\; \sum\limits_{n=1}^\infty
  \; \frac{n}{R}\;
  (\overline{N^m_{l,r}}^{(n)}{N_{l,r}}^{(n)}
  +\overline{N_{l,r}}^{(n)}{N^m_{l,r}}^{(n)}). 
\label{KKn}
\end{eqnarray}
The 
 Dirac mass terms between the bulk
 fields $N$'s and 
 the brane-localized lepton doublets $L$'s
 are given by
\begin{eqnarray}
{\cal L}_{\rm 4Dm}&=&
\frac{1}{\sqrt{M_*}}\sum\limits_{n=0}^\infty
(y^{\nu,L}_{ij} \overline{L_L}_i \tilde{H}_L N_{rj}^{(n)} +
 y^{\nu,R}_{ij} \overline{L_R}_i \tilde{H}_R N_{lj}^{(n)})\delta(y) 
+{\rm h.c.},
\label{4Dm}
\end{eqnarray}
where 
 $M_*$ is the 5D Planck scale.
Then the 4D neutrino mass
 matrix 
 is given by 
\begin{eqnarray}
&& \hspace{-5mm}{\cal L}_{m_\nu}  =  
 (\hspace{3mm} \overline{\nu_L},\hspace{4mm}\overline{N^m_r}^{(1)},
  \hspace{2mm}\overline{N^m_r}^{(2)},
  \cdots \hspace{1mm} | \hspace{5mm}
  \overline{N_l}^{(0)}, \hspace{5mm} \overline{N_l}^{(1)}, 
  \hspace{5mm} \overline{N_l}^{(2)}, \hspace{5mm} \cdots ) \nonumber \\
&&\times
\left(\begin{array}{ccccc|cccc}
  0 & 0& 0 &    \cdots& & 
 \frac{m_L^\nu}{\sqrt{2\pi RM_*}} & 
 \frac{m_L^\nu}{\sqrt{\pi RM_*}} & \frac{m_L^\nu}{\sqrt{\pi RM_*}} 
 & \cdots   \\
    & M_{\cal N} &  &  & &  &\frac{1}{R} & &  \\
    &  & M_{\cal N} &  & &  & &\frac{2}{R} &   \\
   \vdots & & &  \ddots & & & &  &\ddots \\ \hline
   \frac{m_R^{\nu T}}{\sqrt{2\pi RM_*}} &  &   & & & M_{\cal N} & & &   \\
  \frac{m_R^{\nu T}}{\sqrt{\pi RM_*}}
   & \frac{1}{R} &  & &  & & M_{\cal N} & &   \\
   \frac{m_R^{\nu T}}{\sqrt{\pi RM_*}} 
   &  & \frac{2}{R} & &  &  & & M_{\cal N} &   \\
   \vdots &  &  & \ddots & &  & &  &\ddots  
\end{array} 
\right)
\left(
\begin{array}{c}
\hspace{-2mm}\nu_R\hspace{-2mm} \\
\hspace{-2mm}N_l^{m(1)}\hspace{-2mm} \\
\vspace{-1mm}
\hspace{-2mm}N_l^{m(2)}\hspace{-2mm} \\
\vspace{-3mm}
\vdots \\
\vspace{-1mm}
 - \\
\hspace{-2mm}{N_r}^{(0)}\hspace{-2mm} \\
\hspace{-2mm}{N_r}^{(1)}\hspace{-2mm} \\
\hspace{-2mm}{N_r}^{(2)}\hspace{-2mm} \\
\vdots \\
\end{array}
\right)\hspace{-1mm},\;\;
\label{7}
\end{eqnarray}
where 
 $m^\nu_{L,R}$ is triangular matrix 
 in the flavor space as Eq.(\ref{triangular}).

Let us 
 pick up the sub-matrix of $n$-mode from Eq.(\ref{7}). 
Since the KK mass, $n/R$, is proportional to 
 the unit matrix in the flavor space,
 all $n$-mode fields 
 are diagonalized simultaneously 
 in the flavor space as
\begin{equation}
\left(
\begin{array}{cc}
M_{\cal N} & \frac{n}{R} \\
\frac{n}{R} & M_{\cal N}  
\end{array}
\right),  
\label{MV22}
\end{equation}
where 
 $M_{\cal N} = {\rm diag} (M_{\nu e}, M_{\nu \mu}, M_{\nu \tau})$. 
Since the inverse mass matrix of Eq.(\ref{MV22})
 is given by  
\begin{equation}
{1 \over M_{{\cal N}k}^2-(\frac{n}{R})^2}
\left(
\begin{array}{cc}
M_{{\cal N}k} & -\frac{n}{R} \\
-\frac{n}{R} & M_{{\cal N}k}
\end{array}
\right)  
\label{inverse}
\end{equation}
($k$: generation index) 
the summation of the infinite numbers
 of ``seesaw'' is calculated as 
\begin{eqnarray}
\label{9}
m^{\nu}_{ij}&=&
-\frac{(m_L^\nu){}_{ik}}{\pi RM_*}\:  \left[ 
\frac{1}{2M_{{\cal N}k}}+
\sum_{n=1}^{\infty}
{M_{{\cal N}k} \over M_{{\cal N}k}^2-(\frac{n}{R})^2}
\right]
 \: (m_R^{\nu T})_{kj} \nonumber \\
&=& -(m_L^\nu)_{ik}\:
 {1 \over 2 M_* \tan[\pi R M_{{\cal N}k}]}\:
  (m_R^{\nu T})_{kj}, 
\end{eqnarray}
when the magnitude of  
 ${\rm min}[M_{{\cal N}k} \pm \frac{n}{R}]$
 is much 
 larger than the Dirac mass scale\footnote{The case of 
 $|M_k| \ll R^{-1}$  
 reproduces 
 the ordinal 4D universal seesaw formula with 
 the volume suppression factor,
 $({2\pi RM_*})^{-1}$.}. 
This means that the 5D universal 
 seesaw formula is obtained by 
 the replacement 
\begin{equation}
\label{seesaw5D3}
{M_k} 
 \rightarrow 
 {2 M_* \tan [\pi R M_k]}
\end{equation}
{}from the 4D formula\cite{haba}\cite{Dienes:1998sb}. 
Other quarks and leptons' 
 mass matrices are 
 obtained in the same way. 
We should notice that 
 the condition 
 $|m_L|(={\cal O}(10^2)\; {\rm GeV}) \ll |m_R| \ll |M_k|, R^{-1}$  
 is needed for the universal seesaw formula
 and $|m_R/(2M_* \tan [\pi RM_k])|$ should be 
 of ${\cal O}(10^{-2})$ for the suitable mass
 scale of the bottom quark and tau lepton. 
They suggest 
\begin{equation}
M_* \sim R^{-1}={\cal O}(10^{18})\; {\rm GeV}, \;\;\;
|m_R^{b,\tau}/M_k|={\cal O}(10^{-2}).  
\label{condition}
\end{equation}
Other quarks and charged lepton masses can be
 reproduced from the hierarchies of $m_{L,R}$. 
We assume $T,\overline{T}$ are not bulk fields 
 and $|m_R^t|$ should be the same order as 
 (or even larger than) 
 $|M_T|$ 
 in order to reproduce the heavy top quark mass
 as the 4D case.

The tangent function is 
 a feature in the 5D seesaw,
 which has the significant effects
 when
 $|M_k| \sim R^{-1}$. 
Especially, 
 when 
 $|M_k|=\frac{1}{2R}$, 
 the light degrees of freedom 
 become massless by 
 the infinite times ``seesaw'', 
 $-\frac{1}{2\pi R M_*} \times$
 $({[\frac{1}{2R}]^{-1}}-{[\frac{1}{2R}]^{-1}}
 +{[\frac{3}{2R}]^{-1}}-{[\frac{3}{2R}]^{-1}}
 +\cdots ) \rightarrow 0$. 
Therefore, we can explain the smallness
 of Dirac neutrino masses by 
 taking bulk neutrino mass 
 $|M_{{\cal N}k}|$ 
 close to $\frac{1}{2R}$.  
In 4D, the 
 smallness of neutrino masses needed 
 an unnatural assumption of extremely  
 huge $|M_{{\cal N}}|$. 
In 5D, tiny neutrino masses can be 
 naturally realized 
 thanks to the tangent 
 function. 
The neutrino masses can be tiny even 
 when $|M_{\cal N}|$ is the same order
 as other bulk masses.
The condition of $|M_{{\cal N}k}|$ 
 being very close to 
 $\frac{1}{2R}$     
 makes the neutrino masses tiny 
 comparing to other quarks and charged 
 leptons' masses.

\subsection{Majorana neutrino case}

Next, let us show the Majorana 
 neutrino case which corresponds to the 4D case of 
 ${\cal L}'\neq 0$ in Eq.(\ref{15}). 
The quarks and charged lepton sectors are the
 completely same as the previous subsection. 
The difference exists only in the 
 neutrino sector. 
We take the bulk neutrino masses as 
\begin{equation}
\label{3.9}
  {\cal L}_{\rm 5DM} \;=\;
  M_{{\cal N}L}{}_{ij} \overline{N^c_L}_i N_{Lj} +
    M_{{\cal N}R}{}_{ij} \overline{N^c_R}_i N_{Rj} +
    M_{{\cal N}}{}_{ij} \overline{N_L}_i N_{Rj} 
                                              + {\rm h.c.} .
\end{equation}
These masses are invariant under the
 $Z_2$ parity.
The 1st and 2nd masses break
 the lepton number.  
By integrating out the 5th dimensional space, 
 the 
 4D effective Lagrangian is obtained, in which 
 Eq.(\ref{3.9}) becomes 
\begin{eqnarray}
 {\cal L}_{\rm 4DM}\, \;&=&\; 
  \sum\limits_{n=0}^\infty 
 (M_{{\cal N}L} \overline{N_{l}^c}^{(n)}{N_{l}}^{(n)}
 +M_{{\cal N}R} \overline{N_{r}^c}^{(n)}{N_{r}}^{(n)}
 +M_{{\cal N}} \overline{N_{l}}^{(n)}{N_{r}}^{(n)})+{\rm h.c.} \\
&-&
  \sum\limits_{n=1}^\infty 
 (M_{{\cal N}L} \overline{N_{l}^{mc}}^{(n)}{N_{l}^m}^{(n)}
 +M_{{\cal N}R} \overline{N_{r}^{mc}}^{(n)}{N_{r}^m}^{(n)}
 -M_{{\cal N}} \overline{N_{l}^m}^{(n)}{N_{r}^m}^{(n)}) 
 +{\rm h.c.}. \nonumber
\end{eqnarray}
KK masses are given by 
 Eq.(\ref{KKn}). 
The 
 Dirac mass terms between the bulk
 fields $N$'s and 
 the brane-localized lepton doublets $L$'s 
 are given by
\begin{eqnarray}
{\cal L}_{\rm 4Dm}&=&
\frac{1}{\sqrt{M_*}}\sum\limits_{n=0}^\infty
(
 y^{u,L}_{ij} \overline{L_L}_i \tilde{H}_L N_{rj}^{(n)} +
 y^{u,R}_{ij} \overline{L_R}_i \tilde{H}_R N_{lj}^{(n)}  \nonumber \\
&&\hspace{2cm} 
 + \hat{y}^{u,L}_{ij} \overline{N_L}_i \tilde{H}_L N_{lj}^{c(n)} +
   \hat{y}^{u,R}_{ij} \overline{N_R}_i \tilde{H}_R N_{rj}^{c(n)} 
)\delta(y) 
+{\rm h.c.}.
\label{4DmN}
\end{eqnarray}
Then the 4D neutrino mass
 matrix 
 is given by 
\begin{eqnarray}
&& \hspace{-10mm}(\hspace{5mm}\overline{\nu_L}, \hspace{9mm}\overline{\nu^c_R}, 
 \hspace{10mm}\overline{N_l}^{(0)}, \hspace{7mm}\overline{N^c_r}^{(0)}, 
 \hspace{5mm}\overline{N_l}^{(n)}, \hspace{3mm} \overline{N_l^{mc}}^{(n)}, 
 \hspace{2mm}\overline{N^c_r}^{(n)}, 
 \hspace{3mm}\overline{N^m_r}^{(n)})\times \nonumber \\
&& \hspace{-11mm}
\left(\hspace{-3mm}
\begin{array}{cccccccc}
0 & 0 & \frac{\hat{m}_L^\nu}{\sqrt{2\pi RM_*}} & 
  \frac{m_L^\nu}{\sqrt{2\pi RM_*}} 
  &\frac{\hat{m}_L^\nu}{\sqrt{\pi RM_*}} & 0 &
  \frac{m_L^\nu}{\sqrt{\pi RM_*}} 
  & 0   \\
0 & 0 & \frac{m_R^\nu}{\sqrt{2\pi RM_*}} & 
  \frac{\hat{m}_R^\nu}{\sqrt{2\pi RM_*}}
  &\frac{m_R^\nu}{\sqrt{\pi RM_*}} & 0 &
  \frac{\hat{m}_R^\nu}{\sqrt{\pi RM_*}} 
  & 0   \\
\frac{\hat{m}_L^{\nu T}}{\sqrt{2\pi RM_*}} & 
  \frac{m_R^{\nu T}}{\sqrt{2\pi RM_*}} 
  & M_{{\cal{N}}L} & M_{{\cal{N}}} & 0 & 0 & 0 & 0  \\
\frac{m_L^{\nu T}}{\sqrt{2\pi RM_*}} & 
\frac{\hat{m}_R^{\nu T}}{\sqrt{2\pi RM_*}} 
  & M_{{\cal{N}}}^T & M_{{\cal{N}}R}  & 0 & 0 & 0 & 0   \\
\frac{\hat{m}_L^{\nu T}}{\sqrt{\pi RM_*}} & 
\frac{{m}_R^{\nu T}}{\sqrt{\pi RM_*}} 
  & 0 & 0 & M_{{\cal{N}}L} & \frac{n}{R} & M_{{\cal{N}}} & 0 \\
0 & 0 & 0 & 0 & \frac{n}{R} & -M_{{\cal{N}}L} & 0 & M_{{\cal{N}}} \\
\frac{{m}_L^{\nu T}}{\sqrt{\pi RM_*}} & 
\frac{\hat{m}_R^{\nu T}}{\sqrt{\pi RM_*}} & 
0 & 0 & M_{{\cal{N}}}^T & 0 & M_{{\cal{N}}R} & \frac{n}{R} \\
0 & 0 & 0 & 0 & 0 & M_{{\cal{N}}}^T & \frac{n}{R} & -M_{{\cal{N}}R} \\
\end{array}
\hspace{-3mm}\right)\hspace{-2mm}
\left(\hspace{-3mm}
\begin{array}{c}
{\nu_L^c} \\
\vspace{1mm}
{\nu_R} \\
\vspace{1mm}
{N_l^c}^{(0)} \\
\vspace{1mm}
{N_r}^{(0)} \\
\vspace{1mm}
{N_l^c}^{(n)} \\
\vspace{1mm}
{N_l^m}^{(n)} \\
\vspace{1mm}
{N_r}^{(n)} \\
\vspace{1mm}
{N_r^{mc(n)}} \\
\end{array}
\hspace{-3mm}\right)\; 
\label{neu5d}
\end{eqnarray}
where $n=1 \sim \infty$.
This is an infinite times infinite matrix. 
In the 4D limit,
 $|M_{{\cal N}L}|, |M_{{\cal N}R}|, |M_{{\cal N}}| \ll R^{-1}$,
 the infinite mass matrix Eq.(\ref{neu5d})
 reduces to 4D mass matrix Eq.(\ref{neu4d}).

Here let us consider some special limits.
The first is $M_{{\cal N}L,R}=0$ and $\hat{m}_{L,R}^\nu=0$ 
 which is the limit of previous subsection, and 
 can not induce seesaw mechanism because 
 it does not break lepton number. 
Next is the limit of $M_{{\cal N}}=0$ and $\hat{m}^\nu_{L,R}=0$. 
This case makes Eq.(\ref{neu5d}) become 
\begin{eqnarray}
&& 
\hspace{-7mm}(\hspace{3mm}\overline{\nu_L}, 
\hspace{9mm}\overline{N^m_r}^{(n)} |
\hspace{3mm}\overline{N^c_r}^{(0)}, 
\hspace{7mm}\overline{N^c_r}^{(n)} \hspace{1mm}|\hspace{-.5mm}| 
\hspace{5mm}\overline{\nu^c_R},
\hspace{8mm}\overline{N_l^{mc}}^{(n)}| 
\hspace{5mm}\overline{N_l}^{(0)}, 
\hspace{5mm}\overline{N_l}^{(n)}  
\hspace{2mm})\times \nonumber \\
&& \hspace{-8mm}
\left(\hspace{-3mm}
\begin{array}{cc|cc||cc|cc}
0 & 0 & \frac{{m}_L^\nu}{\sqrt{2\pi RM_*}} 
  &\frac{{m}_L^\nu}{\sqrt{\pi RM_*}} & 0 & 0 & 0 & 0 \\
0 & -M_{{\cal{N}}R} & 0 & \frac{n}{R} & 0 & 0 & 0 & 0 \\ \hline
\frac{{m}_L^{\nu T}}{\sqrt{2\pi RM_*}} & 0 & M_{{\cal{N}}R} 
  & 0 & 0 & 0 & 0 & 0 \\
\frac{{m}_L^{\nu T}}{\sqrt{\pi RM_*}}  & \frac{n}{R} & 0 & M_{{\cal{N}}R} 
& 0 & 0 & 0 & 0  \\ \hline\hline
0 & 0 & 0 & 0 & 
0 & 0 & \frac{m_R^\nu}{\sqrt{2\pi RM_*}} 
  &\frac{m_R^\nu}{\sqrt{\pi RM_*}} \\
0 & 0 & 0 & 0 & 
0 & -M_{{\cal{N}}L} & 0 & \frac{n}{R} \\ \hline
0 & 0 & 0 & 0 & 
\frac{m_R^{\nu T}}{\sqrt{2\pi RM_*}} & 0 & M_{{\cal{N}}L} & 0 \\
0 & 0 & 0 & 0 & 
\frac{m_R^{\nu T}}{\sqrt{\pi RM_*}}  & \frac{n}{R} & 0 & M_{{\cal{N}}L} \\ 
\end{array}
\right)\hspace{-2mm}
\left(\hspace{-3mm}
\begin{array}{c}
{\nu_L^c} \\
\vspace{1mm}
{N_r^{mc(n)}} 
\vspace{-2.5mm} \\ 
- 
\vspace{-1mm} \\
{N_r}^{(0)} \\
\vspace{1mm}
{N_r}^{(n)} 
\vspace{-2mm}\\
- \vspace{-4.5mm}\\
- \vspace{-3.5mm}\\
\vspace{1mm}
{\nu_R} \\
\vspace{1mm}
{N_l^m}^{(n)}  
\vspace{-2mm} \\ 
- 
\vspace{-1mm} \\
{N_l^c}^{(0)} \\
\vspace{1mm}
{N_l^c}^{(n)} 
\vspace{1mm}
\end{array}
\hspace{-3mm}\right). \nonumber
\label{neu5d1}
\end{eqnarray}
Two same structures 
 appear in upper-left 
 and downer-right parts in this matrix. 
This means 
 two 
 $3\times3$ mass matrices are obtained 
 through
 the seesaw mechanism,
 when $|m_{L,R}^\nu | \ll |M_{{\cal N}L,R}|, R^{-1}$. 
They are given by \cite{haba}\cite{Lukas:2000rg}-\cite{later}
\begin{eqnarray}
\label{9}
m^{\nu_l}_{ij} 
&\simeq& -(m_L^\nu)_{ik}\:
 {1 \over 2 M_* \tanh[\pi R M_{{\cal N}R}{}_k]}\:(m_L^{\nu T})_{kj}, \\
m^{\nu_r}_{ij} 
&\simeq& -(m_R^\nu)_{ik}\:
 {1 \over 2 M_* \tanh[\pi R M_{{\cal N}L}{}_k]}\:(m_R^{\nu T})_{kj} .
\label{10}
\end{eqnarray}
When $|m_R^{\nu}/M_k|={\cal O}(10^{-2})$ is
 satisfied as Eq.(\ref{condition}), 
 the bulk mass $|M_{\cal N}{}_{Rk}|$ should be of
 ${\cal O}(10^{14})$ GeV for the suitable 
 mass scale 
 $|m^{\nu_l}|={\cal O}(10^{-1})$ eV of 
 the three generation active 
 neutrinos. 
It is natural to consider 
 $|M_{\cal N}{}_{Rk}| \sim |M_{\cal N}{}_{Lk}|$, 
 which suggests 
 $|m^{\nu_r}|={\cal O}(10^{10})$ GeV from 
 Eq.(\ref{condition}). 
The limit, 
 $M_{{\cal N}}=0$ and 
 ${m}_{L,R}=0$, 
 changes Eqs.(\ref{9}) and (\ref{10}) as 
\begin{eqnarray}
\label{9p}
m^{\nu_l}_{ij} 
&\simeq& -(m_L^\nu)_{ik}\:
 {1 \over 2 M_* \tanh[\pi R M_{{\cal N}L}{}_k]}\:(m_L^{\nu T})_{kj}, \\
m^{\nu_r}_{ij} 
&\simeq& -(m_R^\nu)_{ik}\:
 {1 \over 2 M_* \tanh[\pi R M_{{\cal N}R}{}_k]}\:(m_R^{\nu T})_{kj} .
\label{10p}
\end{eqnarray}
Both limits satisfy 
 the 4D relation 
\begin{equation}
|m^{\nu_l}| |m^{\nu_r}| \sim |m^\tau|{}^2 
\end{equation}
when $|M_{{\cal N}L,R}|\sim |M_{{\cal E}}|$.



\section{Summary}

We have shown the universal seesaw mechanism 
 in the flat extra dimension setup. 
The 5D universal seesaw has been demonstrated 
 in the framework of the left-right symmetric model. 
We introduce  
 extra 
 vector-like isosinglet fields 
 in the 5D bulk. 
The universal seesaw formula is easily obtained as
 a replacement of the vector-like mass in 4D case $M_i$ to
 $2 M_* \tan [\pi RM_i]$ ($M_*$: 5D Planck scale,
 $M_i$: vector-like bulk mass, and 
 $R$: compactification radius). 
As for the neutrinos, 
 the smallness of Dirac neutrino masses 
 can be naturally explained in the 5D setup. 
Only Dirac neutrino masses can be very small 
 when the bulk neutrino masses are close to a 
 half of
 compactification scale. 
We have also shown the Majorana neutrino case, where
 5D seesaw mechanism works. 
For the large top mass, 
 the vector-like top quarks should be
 localized on the 4D brane.

We can also consider 
 the universal seesaw
 mechanism in the 
 frame work of $SU(5)$ GUT. 
When heavy three generation vector-like 
 ${\bf 10}$ and
 ${\bf 1}$
 fields and singlet Higgs field are
 introduced in the bulk,
 the universal seesaw mechanism 
 works in the 5D $SU(5)$ GUT\cite{SU5}. 
The 5D universal seesaw calculations are
 easily achieved by the replacement of 
 $\mu_{10} \rightarrow 
 \tanh[2\pi R \mu_{10}]$
 in Eqs.(3.6)-(3.8) and 
 $\langle 1'_H \rangle \rightarrow 
 \tanh[2\pi R \langle 1'_H\rangle]$
 in Eq.(3.9) 
 of Ref.\cite{SU5}.

\vskip 1cm

\leftline{\bf Acknowledgments}
N.H. is supported by the Alexander von Humboldt Foundation,
 and would like to thank 
 T. Ota 
 for fruitful discussions.

\vspace{.5cm}


\end{document}